\documentclass{llncs}
\def\Q{{\bf Q}}
\def\R{{\bf R}}
\def\Res{\mathop{\rm Res}\nolimits}
\usepackage{color}

\begin{document}

\title{The Potential and Challenges of CAD with Equational Constraints for SC-Square}
\titlerunning{The Potential and Challenges of CAD with Equational Constraints for SC-Square}  
%
\author{James H. Davenport\inst{1} \and Matthew England\inst{2}}
\authorrunning{J.H. Davenport and M. England} 
%
%
\institute{
Department of Computer Science, 
University of Bath, Bath, UK \\
\email{J.H.Davenport@bath.ac.uk}
\and
Faculty of Engineering, Environment and Computing, \\
Coventry University, Coventry, UK \\
\email{Matthew.England@coventry.ac.uk}
}

\maketitle              

\begin{abstract}
Cylindrical algebraic decomposition (CAD) is a core algorithm within Symbolic Computation, particularly for quantifier elimination over the reals and polynomial systems solving more generally.  It is now finding increased application as a decision procedure for Satisfiability Modulo Theories (SMT) solvers when working with non-linear real arithmetic.  
We discuss the potentials from increased focus on the logical structure of the input brought by the SMT applications and \textsf{SC}$^2$ project, particularly the presence of equational constraints.  We also highlight the challenges for exploiting these: primitivity restrictions, well-orientedness questions, and the prospect of incrementality.
\end{abstract}

\section{Introduction}

\subsection{Cylindrical Algebraic Decomposition}

The original aim of Cylindrical Algebraic Decomposition (CAD), as introduced by \cite{Collins1975}, was \emph{Quantifier Elimination} (QE).  
More precisely, given
\begin{equation}\label{eq:Phi}
Q_{k+1}x_{k+1}\ldots Q_nx_n\Phi(x_1,\ldots,x_n)
\end{equation}
where $Q_i\in\{\forall,\exists\}$ and $\Phi$ is a Tarski Formula;
produce an equivalent formula, $\Psi(x_1,\ldots,x_k)$ which is quantifier free.  Here a \emph{Tarski Formula} is a Boolean combination of predicates $f_j \, \sigma_j , 0$ with $\sigma_j \in \{=,\ne,>,\ge,<,\le\}, f_j\in\Q[x_1,\ldots,x_n]$.

The CAD of \cite{Collins1975} was a major breakthrough, with a running time ``merely'' doubly-exponential in $n$, as opposed to previous methods \cite{Tarski1951}.

The methodology of Collins' CAD is broadly as follows:
\begin{enumerate}
\item Retaining from (\ref{eq:Phi}) only the $f_j$ (call this set $S_n$) and the order of the $x_i$, compute a CAD of $\R^n$, sign-invariant for the $f_i$.\label{Step:CAD}
\begin{enumerate}
\item Repeatedly project $S_\ell\subset\Q[x_1,\ldots,x_{\ell}]$ to $S_{\ell-1}:=P_C(S_{\ell})\subset\Q[x_1,\ldots,\allowbreak x_{\ell-1}]$ where $P_C$ is Collins' projection operator. 
\item Isolate real roots of $S_1$ to produce a CAD of $\R^1$ sign-invariant for $S_1$.
\item Repeatedly lift the decomposition of $\R^{\ell-1}$ to one of $\R^\ell$, sign-invariant for $S_\ell$.  
To lift over a cell in $\R^{\ell-1}$ we substitute a sample point of the cell into $S_{\ell}$; perform univariate root isolation and decompose accordingly.  $P_C$ is chosen so that the sample point is representative of the whole cell. 
\end{enumerate}
\item Using the $Q_i$ and $\Phi$ identify cells of the induced CAD of $\R^k$ true for (\ref{eq:Phi}).
\item Deduce $\Psi$.
\end{enumerate}
There have been many improvements since \cite{Collins1975}: we quote only two here, referring to \cite{Bradfordetal2016a} for a more detailed summary.

\textbf{\cite{McCallum1984}:}  This replaced the operator $P_C$ by a much smaller operator $P_M$, simultaneously replacing ``sign-invariant'' by ``order-invariant'' in Step \ref{Step:CAD}. 
However, the system has to be ``well-oriented'', which can only be seen with hindsight, when a lack of it manifests itself by a polynomial being nullified, i.e. vanishing entirely, over a cell of dimension $>0$.

\textbf{\cite{Lazard1994}:}  This replaced the operator $P_M$ by a slightly smaller operator $P_L$ and significantly modified the lifting procedure, simultaneously replacing ``order-invariant'' by what is now called ``Lazard-valuation-invariant'' in Step \ref{Step:CAD}.  
A gap in the proof of \cite{Lazard1994} was soon spotted. It was rectified recently by \cite{McCallumHong2016a}, but using the technology of order-invariant and under the well-oriented restriction.
A complete resolution, in terms of Lazard invariance, has been presented in the preprint \cite{McCallumetal2017a}.

\subsection{New Applications: \textsf{SC}$^2$}

The authors are involved in the EU Project \textsf{SC}$^2$ which aims to forge interaction between the communities of \textbf{S}ymbolic \textbf{C}omputation and \textbf{S}atisfiability \textbf{C}hecking \cite{SC2}.  CAD and QE are traditionally found in the former but recently the technology behind them have been applied in Satisfiability Modulo Theory (SMT) solvers \cite[for example]{JdM12} where the problem is usually not to perform full QE but to test satisfiability, finding either a witness point or (minimal) proof of unsatisfiability.  Such solvers are used routinely in industries such as software verification.  The problem sets are different to those typical in CAD: often lower degree polynomials but far more of them and in more variables.  Viewed from Satisfiability Checking the CAD procedure outlined above is curious, particular in its discarding of the logical structure in Step \ref{Step:CAD}.

\section{Potentials}

\subsection{Equational Constraint}

The fact that the $\sigma_j$ and $\Phi$ are essentially ignored in Step \ref{Step:CAD} was noticed in \cite{Collins1998}, at least for the special case 
\begin{equation}\label{eq:Phi1}
\Phi(x_1,\ldots,x_n) \equiv F_1(x_1,\ldots,x_n)=0\land\Phi'(x_1,\ldots,x_n)
\end{equation}
(where $F_1$ depends non-trivially on $x_n$ and is primitive): intuitively the key idea is that we do not care about the polynomials in $\Phi'$ away from $F_1=0$.  We refer to $F_1=0$ as an \emph{equational constraint} (more generally, an equation implied by the formula). This was formalised in \cite{McCallum1999a}. The key result there is the following.

\def\foo{\cite[Theorem 2.2]{McCallum1999a}}
\begin{theorem}[\foo]
Let $r > 2$, let $f(x_1,\ldots,x_r)$ and $g(_1,\ldots,x_r)$ be real polynomials of positive degrees in the main variable $x_r$, let $R(x_1,\ldots,x_{r-1})$ be the resultant of $f$ and $g$, and suppose that $R\ne 0$. Let $S$ be a connected subset of $\R^{r-1}$ on which $f$ is delineable and in which $R$ is order-invariant. Then $g$ is sign-invariant in each section of $f$ over $S$.
\end{theorem}
In the context of (\ref{eq:Phi1}) this justifies replacing $P_M(S_r)$ by the reduced projection operator
\begin{equation}
P_M(F_1;S_r):=P_M(\{F_1\})\cup\{\Res_{x_r}(F_1,f_i): f_i \in \Phi'\},
\label{eq:ECProj}
\end{equation}
at least for the first projection.
If $S_r$ has $n$ polynomials of degree $d$,  $P_M(S_r)$  has $\frac12n(n+1)$ polynomials of degree $O(d^2)$ whereas $P_M(F;S_r)$ has $n$ such.  

\subsection{Multiple Equational Constrains}

If there are multiple equational constraints then it is possible to use a variant (slightly enlarged) of the reduced operator (\ref{eq:ECProj}) for projections beyond the first.  The idea is to \emph{propagate} the constraints by noticing their resultant is also implied by the formula but does not contain the main variable \cite{McCallum2001}.  

More recently, in \cite{EBD15} the present authors identified savings in the lifting phase: the fact that Theorem 1 provides not just delineability but sign-invariance for $g$ means there is no need to isolate and decompose with respect to the real roots of $g$.  This, combined with the use of Gr\"obner Basis technology to control the degree growth of projection polynomials allowed us to present an improved complexity analysis of CAD with multiple equational constraints in \cite{ED16}.  Broadly speaking, we decrease the double exponent by one for each equational constraint. 

\subsection{Equational Constraints of Sub-formulae}

If instead of (\ref{eq:Phi1}), our problem has the form
\begin{equation}\label{eq:Phi2}
\Phi(x_1,\ldots,x_n) \equiv \left(f_1=0\land\Phi_1\right)\lor\left(f_2=0\land\Phi_2\right)\lor\cdots,
\end{equation}
then we can write it in the form (\ref{eq:Phi1}) by letting $F_1=\prod f_i$. However, as was observed in \cite{Bradfordetal2013b}, we can do better by analysing the inter-dependencies in (\ref{eq:Phi2}) more carefully, building a truth-table invariant CAD (TTICAD) for the collection of sub-formulae.   Intuitively the key idea is that we do not care about the polynomials in $\Phi_i$ outside $f_i=0$. TTICAD was expanded in \cite{Bradfordetal2016a} to the case where not every disjunct has an equation (so  there is no overall equational constraint for $\Phi$).

\section{Challenges}

Section 2 identifies a wealth of technology for making greater use of the logical structure of the CAD input.  However, there are a number of challenges.

\subsection{Need for primitivity}

All the theory of reduced projection operators requires that the constraint be primitive.  No technology currently exists (beyond reverting to sign-invariance) for the non-primitive case (although ideas were sketched in \cite{EBD15}).  Note that the restriction is not just on the input but also constraints found through propagation.  In \cite{DE16} the Davenport-Heinz examples \cite{DH88} used to demonstrate the doubly exponential complexity of CAD were shown to lack primitivity, showing that the non-primitive case is genuinely more difficult.

\subsection{Well-orientedness}

All the existing theory of reduced projection operators rests on the mathematics of order-invariance developed for $P_M$.  The reduced operators not only require this condition of $P_M$ but actually extend it (they are less complete).  The lack of this condition is only discovered at the end of CAD (when we lift with respect to the offending polynomials).  
For traditional CAD this means a large waste of resources starting the calculation again. 

As described in Section 1 there is a new sign-invariant projection operator $P_L$ which achieves the savings of $P_M$ without sacrificing completeness.  It may be possible to expand this to a family of reduced operators, but this requires development of the corresponding Lazard valuation invariance theory.

\subsection{Incremental CAD}

A key requirement for the effective use of CAD by SMT-solvers is that the CAD be incremental: that polynomials can be added and removed to the input with the data structures of the CAD edited rather than recalculated.  Such incremental CAD algorithms are now under development by the \textsf{SC}$^2$ project \cite{SC2}.

These could offer a partial solution to the difficulties of well-orientedness.  I.e. if a particular operator is found to not be well-oriented at the end of a CAD calculation the next step would be to revert to a less efficient operator which is usually a superset of the original one.  Hence we could edit the existing decomposition to take into account these additional polynomials.

However, the use of CAD with equational constraints incrementally may exhibit strange behaviour in the SMT context.  For example, removing a constraint that was equational could actually grow the output CAD since it necessitates the use of a larger projection operator.  Correspondingly, adding an equational constraint could allow a smaller operator and shrink the output.  It is not clear how SMT solvers heuristics should be adapted to handle these possibilities.


\subsubsection*{Acknowledgements:}

The authors are supported by the European Union's Horizon 2020 research and innovation programme under grant agreement No H2020-FETOPEN-2015-CSA 712689 (\textsf{SC}$^2$).

\newcommand{\etalchar}[1]{$^{#1}$}


\begin{thebibliography}{BDE{\etalchar{+}}16}

\bibitem[SC2]{SC2}
E.~{\'A}brah{\'a}m, J.~Abbott, B.~Becker, A.M. Bigatti, M.~Brain,
  B.~Buchberger, A.~Cimatti, J.H. Davenport, M.~England, P.~Fontaine,
  S.~Forrest, A.~Griggio, D.~Kroening, W.M. Seiler, and T.~Sturm.
\newblock $\mathsf{SC}^2$: Satisfiability checking meets symbolic computation.
\newblock In M.~Kohlhase et al. (eds), {\em Intelligent Computer Mathematics}, LNCS 9791, pages 28--43. 
Springer, 2016.

\bibitem[BDE{\etalchar{+}}13]{Bradfordetal2013b}
R.J. Bradford, J.H. Davenport, M.~England, S.~McCallum, and D.J. Wilson.
\newblock {Cylindrical Algebraic Decompositions for Boolean Combinations}.
\newblock In {\em Proc. ISSAC 2013}, pages 125--132, 2013.

\bibitem[BDE{\etalchar{+}}16]{Bradfordetal2016a}
R.J. Bradford, J.H. Davenport, M.~England, S.~McCallum, and D.J. Wilson.
\newblock {Truth table invariant cylindrical algebraic decomposition}.
\newblock {\em J. Symbolic Computation}, 76:1--35, 2016.

\bibitem[Col75]{Collins1975}
G.E. Collins.
\newblock {Quantifier Elimination for Real Closed Fields by Cylindrical
  Algebraic Decomposition}.
\newblock In {\em Proc. 2nd. GI Conference Automata Theory \& Formal
  Languages}, pages 134--183, 1975.

\bibitem[Col98]{Collins1998}
G.E. Collins.
\newblock {Quantifier elimination by cylindrical algebraic decomposition ---
  twenty years of progess}.
\newblock In B.F. Caviness and J.R. Johnson, editors, {\em Quantifier
  Elimination and Cylindrical Algebraic Decomposition}, pages 8--23. Springer Verlag, Wien, 1998.

\bibitem[DE16]{DE16}
J.H. Davenport and M.~England.
\newblock Need Polynomial Systems be Doubly-exponential?
\newblock Proc. ICMS 2016, Lecture Notes in Computer Science 9725, Springer, 2016, pp.  157-164.

\bibitem[DH88]{DH88}
J.H. Davenport and J.~Heintz.
\newblock Real quantifier elimination is doubly exponential.
\newblock {\em Journal of Symbolic Computation}, 5(1-2):29--35, 1988.

\bibitem[EBD15]{EBD15}
M.~England, R.~Bradford, and J.H. Davenport.
\newblock Improving the use of equational constraints in cylindrical algebraic   decomposition.
\newblock In Proc. ISSAC '15, pages 165--172. ACM, 2015.

\bibitem[ED16]{ED16}
M.~England and J.H. Davenport.
\newblock The complexity of cylindrical algebraic decomposition with respect to polynomial degree.
\newblock In V.P. Gerdt et al. (eds), {\em Computer Algebra in Scientific Computing}, LNCS 9890, pages 172--192. Springer International Publishing, 2016.

\bibitem[JdM12]{JdM12}
D.~Jovanovic and L.~de~Moura.
\newblock Solving non-linear arithmetic.
\newblock In B.~Gramlich et al. (eds), Proc. IJCAR '12, LNCS 7364, pages 339--354. Springer, 2012.

\bibitem[Laz94]{Lazard1994}
D.~Lazard.
\newblock {An Improved Projection Operator for Cylindrical Algebraic
  Decomposition}.
\newblock In {\em Proc. Algebraic Geometry and its Applications}, 1994.

\bibitem[McC84]{McCallum1984}
S.~McCallum.
\newblock {\em {An Improved Projection Operation for Cylindrical Algebraic
  Decomposition}}.
\newblock PhD thesis, University of Wisconsin-Madison Computer Science, 1984.

\bibitem[McC99]{McCallum1999a}
S.~McCallum.
\newblock {On Projection in CAD-Based Quantifier Elimination with Equational
  Constraints}.
\newblock In {\em Proc. ISSAC '99}, pages 145--149,
  1999.
  
\bibitem[McC01]{McCallum2001}
S.~McCallum.
\newblock On propagation of equational constraints in {CAD}-based quantifier elimination.
\newblock In Proc. ISSAC 2001, pages 223--231. ACM, 2001.  

\bibitem[MH16]{McCallumHong2016a}
S.~McCallum and H.~Hong.
\newblock {On Using Lazard's Projection in CAD Construction}.
\newblock {\em J. Symbolic Computation}, 72:65--81, 2016.

\bibitem[MPP17]{McCallumetal2017a}
S. McCallum, A.~Parusinski and L.~Paunescu \\
\newblock Arxiv \url{https://arxiv.org/abs/1607.00264v2},
\newblock 2017


\bibitem[Tar51]{Tarski1951}
A.~Tarski.
\newblock {\em {A Decision Method for Elementary Algebra and Geometry}}.
\newblock 2nd ed., Univ. Cal. Press. Reprinted in {\em Quantifier Elimination
  and Cylindrical Algebraic Decomposition} (ed. B.F. Caviness \& J.R. Johnson),
  Springer-Verlag, Wein-New York, 1998, pp. 24--84., 1951.

\end{thebibliography}
\end{document}